\documentclass[runningheads]{llncs}
\usepackage[T1]{fontenc}
\usepackage{graphicx}
\usepackage{booktabs}
\usepackage{amsmath}
\usepackage[misc]{ifsym}

\usepackage{amssymb}
\usepackage{array}
\usepackage{mwe}
\usepackage[algo2e, ruled,linesnumbered]{algorithm2e}
\usepackage{caption}
\usepackage{multirow}
\usepackage{subcaption} % 支持子图

\begin{document}

\title{Federated Mixture-of-Expert for Non-Overlapped Cross-Domain Sequential Recommendation}

\titlerunning{FMoE-CDSR}
% 

%
% \author{Author information scrubbed for double-blind reviewing}
% 
% 
\author{Yu Liu\inst{1}$^\S$ \and
Hanbin Jiang\inst{1}$^\S$ \and
Lei Zhu\inst{1}$^\S$ \and
Yu Zhang\inst{1}$^\S$ \and
Yuqi Mao\inst{1} \and \\
Jiangxia Cao\inst{2} (\Letter)\and
Shuchao Pang\inst{1}  (\Letter) \thanks{$\S$ Equal contributions, \Letter\   Corresponding authors.}
}
\authorrunning{Yu Liu, Hanbin Jiang, Lei Zhu, Yu Zhang, Yuqi Mao, Jiangxia Cao, Shuchao Pang}
\tocauthor{Yu Liu, Hanbin Jiang, Lei Zhu, Yu Zhang, Yuqi Mao, Jiangxia Cao, Shuchao Pang}
\institute{Nanjing University of Science and Technology \and Kuaishou Technology\\
\email{\{lovekikyo,jianghanbin,zhulei,zy922128980209,yuqimao,pangshuchao\}@njust.edu.cn},
\email{caojiangxia@kuaishou.com}}

\maketitle           

\begin{abstract}
   In the real world, users always have multiple interests while surfing different services to enrich their daily lives, e.g., watching hot short videos/live streamings.
    To describe user interests precisely for a better user experience, the recent literature proposes cross-domain techniques by transferring the other related services (a.k.a.
     domain) knowledge to enhance the accuracy of target service prediction.
   In practice, naive cross-domain techniques typically require there exist some overlapped users, and sharing overall information across domains, including user historical logs, user/item embeddings, and model parameter checkpoints.
    Nevertheless, other domain's user-side historical logs and embeddings are not always available in real-world RecSys designing, since users may be totally non-overlapped across domains, or the privacy-preserving policy limits the personalized information sharing across domains.
    Thereby, a challenging but valuable problem is raised: \textbf{How to empower target domain prediction accuracy by utilizing the other domain model parameters checkpoints only?}
    To answer the question, we propose the FMoE-CDSR, which explores the non-overlapped cross-domain sequential recommendation scenario from the federated learning perspective.
    Specifically, we devise a novel federated mixture-of-expert paradigm to maximize other domain expert's effectiveness on the target domain.
    Experimental results demonstrate that our method significantly enhances recommendation performance across multiple real-world scenarios.
    Besides, we also conduct comprehensive ablation studies and detailed analyses to investigate the effectiveness of our model components.

\end{abstract}

\section{Introduction}
With the rapid development of the internet and digital technologies, various types of information on the internet have experienced explosive growth. 
To distribute them, the recommendation system (RS) became a more and more important way to send personalized content to our users to meet multiple demands in their daily lives, e.g., listening to fresh news~\cite{chen2024multi}, buying some goods~\cite{cheng2025choruscvr}, and watching hot short videos/live streamings~\cite{cao2024moment}.
To provide a better recommendation experience for users, the collaborative filter-based RS is needed to utilize the life-long users' history logs to learn his/her interests point to make an appropriate prediction.
Therefore, as a promising technique for transferring the other services (a.k.a. domain) knowledge to the target domain, the cross-domain technique~\cite{zang2022survey,cao2022contrastive} has been a hot topic in recent years, to enhance target services recommendation accuracy.

In a broad sense, the cross-domain recommendation has an assumption that there exists a certain number of overlapped users that act as bridges to connect different domain interactions~\cite{cao2022disencdr,cao2023towards}.
Based on these overlapped users, the naive cross-domain techniques typically required sharing the overall information across domains, including the (i) user historical logs, (ii) user/item embeddings, and (iii) model parameters checkpoints.
However, such limitations sometimes do not hold true in the real world: (1) it is difficult to align users across domains, e.g., TikTok and Amazon. (2) sharing user-side historical logs or embeddings poses a risk of privacy-preserving exposure~\cite{kairouz2021advances}.
Compared to these risky to obtained overlapped users and embeddings, the trained model parameters checkpoints of different domains are safer to share with other domains.
Therefore, in this paper, we focus on exploring a challenging but valuable problem: \textbf{How to empower target domain prediction accuracy by utilizing the other domain model parameters checkpoints only?}

\begin{figure*}[t]
    \begin{center}
        \includegraphics[width=1.0\textwidth]{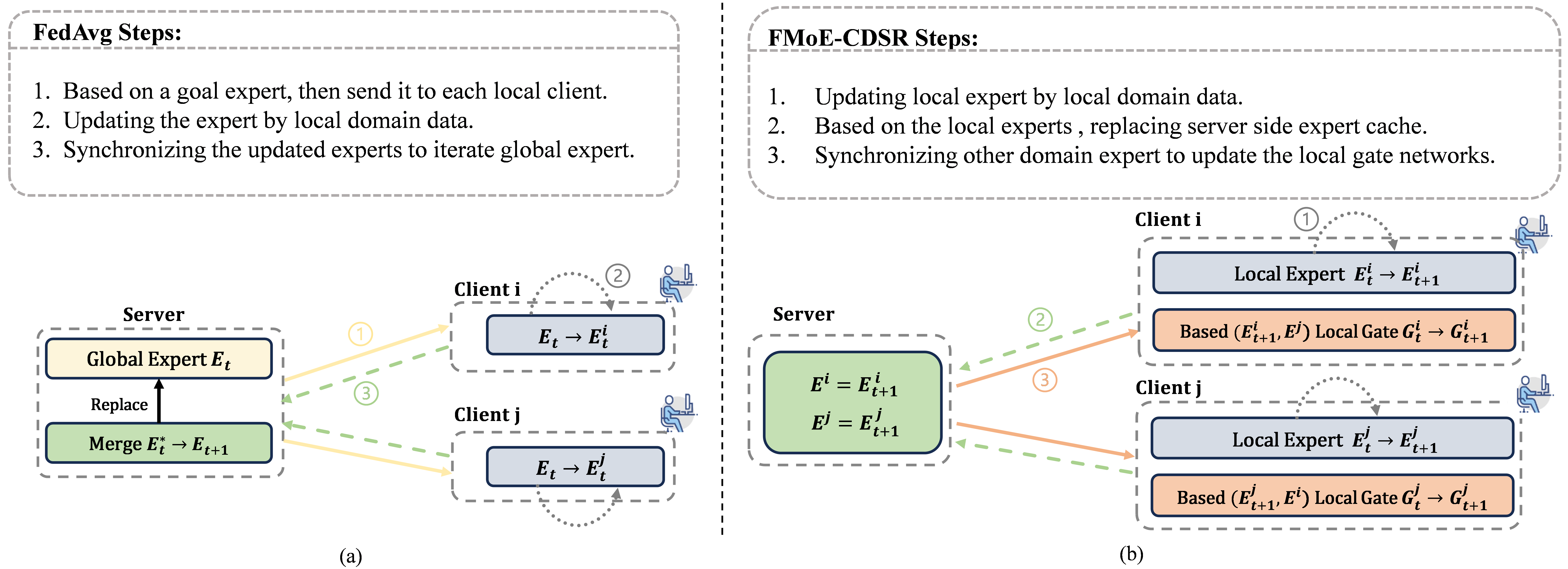}
        \caption{Model steps of FedAvg and FMoE-CDSR.}
        \label{intro}
    \end{center}
\end{figure*}

Actually, the above problem is closely related to Federated Learning, which trains models in each client (e.g., domain) and only shares model updates instead of original data, thereby protecting privacy and reducing data transmission.
Following the Federated Learning paradigm, the FedCDR~\cite{fedcdr}, PriCDR~\cite{PriCDR}, P2FCDR~\cite{P2FCDR}, FedDCSR~\cite{zhang2024feddcsr}, and other cross-domain efforts are proposed, however, they are still made some strong assumptions: (i) the first branch works assume the users are overlapped while focus on transferring users embeddings by adding the differential privacy noise, (ii) the second branch works aims at transferring model parameters by utilizing a simple model parameter averaging mechanism (FedAvg~\cite{fedavg}), as shown in Figure~\ref{intro}(a).
Specifically, the FedAvg is a wide-used technique to communicate different domain models, which includes three major steps in one iteration: (1) initialize a global shared model at the server side and then send the model to each local client, (2) using in-domain data to update each domain local client, (3) synchronize the local client model's full parameters to the server side, and then average the parameters to update global model.

Although the FedAvg~\cite{fedavg} based cross-domain methods achieve promising results to some extent, however,  they still have the following drawbacks: (i) the pattern of user behavior characteristics in different domains is highly different, the simple FedAvg mechanism has significant information loss to effectively reflect the unique knowledge of each domain, (ii) those methods suppose there exists a same model architecture across domain, which limits the flexibility of model designing in different domain.
In this paper, inspired by the success of Mixture-of-Experts (MoE~\cite{ma2018modeling,wang2024home}) in the recommendation, we propose a novel federated learning framework for a non-overlapped cross-domain sequential recommendation, termed as \textbf{FMoE-CDSR}.
Specifically, our FMoE-CDSR consists of the following designing insights (as shown in Figure~\ref{intro}(b)): (1) First, we regard the model of each domain (e.g. client) as an expert, which contains the recommendation knowledge of the corresponding domain. (2) Based on different domain's local models, we next synchronize other domain experts to the target domain and assign an additional parameters to adapt them to target domain prediction. (3) We further introduce MoE mechanism to select the in-domain experts and the synchronized experts adaptively to aggregate different expert knowledge for more comprehensive recommendations.

In summary, our contributions are as follows:
\begin{itemize}
    \item We propose a novel federated learning paradigm for cross-domain sequential recommendation called FMoE-CDSR, which explores a more generalized cross-domain setting that non-overlapped users bridge domains.
    \item We introduce a flexible federated mixture-of-experts mechanism, which effectively leverages model parameters from multiple other domains without any architecture limitation.
    \item We conduct extensive experiments on three real-world CDR datasets to verify our model performance. Further, we conduct detailed ablation analyses to investigate the effectiveness of our model learning paradigm.
    % \footnote{Code will be available at \url{Github}}.
\end{itemize}

\section{Preliminary}

\subsection{Problem Definition}
Assuming there exists 3 domains $i,j,k$, let the $\mathcal{D}^k=\{\big((s_1^k, s_2^k,\dots,s_T^k), s_{T+1}^k\big)_u\}$ denotes domain $k$'s data samples, where the user $u$ from user set $\mathcal{U}^k$.
For each user $u$'s data sample, the $(s_1^k, s_2^k,\dots,s_T^k)$ indicates the historical interacted item ID sequence with length $T$, and the $s_{T+1}^k$ denotes the ground truth next item ID, where each item $s^k$ belongs to domain $k$-th item set $\mathcal{S}^k$.
Besides, we also introduce several directed transportation adjacent matrices to compress domain's item-item relationship, e.g., $\mathbf{A}^k\in \{0,1\}^{|\mathcal{S}^k|\times|\mathcal{S}^k|}$, where each  element $\mathbf{A}^k_{n,m} = 1$ denotes the $s_m^k$ one of the next item of $s_n^k$, and $\mathbf{A}^k_{n,m} = 0$ otherwise.

\section{Proposed Model}
\subsection{Overview of FMoE-CDSR}
The proposed FMoE-CDSR operates on a client-server federated learning framework, which lets each client train models specially and locally without leaking data to other clients, only the local model parameters are uploaded to a central server. The model for each client is divided into the local expert and the global experts synchronized from other domains. During each training round, clients synchronize with the server by downloading updated global expert parameters. Upon completing local training, clients exclusively transmit their local expert parameters to the central server, ensuring privacy preservation in cross-domain knowledge aggregation.

\subsection{Embedding Layer}
In embedding LookUp stage, we introduce several parameter matrices to support the trainable local domain expert network data fitting and un-trainable global synchronized expert networks domain adaptation\footnote{\textbf{The local means target domain, and global means other domain.}}.
For example, in domain $k$, we initialize one \textbf{local matrix} $\mathbf{S}^k \in \mathbb{R}^{|\mathcal{S}^k|\times d}$, and two \textbf{global matrices} $\mathbf{S}^{k,i},\mathbf{S}^{k,j} \in \mathbb{R}^{|\mathcal{S}^k|\times d}$, where $d$ means the representation dimension.
Based on them, for arbitrary item ID $s^k \in \mathcal{S}^k$, we could utilize the $\mathbf{s}^k, \mathbf{s}^{k,i}, \mathbf{s}^{k,j}$ to represent its embedding.
Additionally, for the sequence \textbf{position encoding}, we introduce the $\mathbf{P}^k, \mathbf{P}^{k,i}, \mathbf{P}^{k,j} \in \mathbb{R}^{T\times d}$ for the causality self-attention modeling.

\subsection{Local Domain Expert Learning}
In this section, we first describe a typical workflow of training local domain expert in sequential recommendation task.
For the encoder part, our federated paradigm actually supports that different domains have their own encoder architecture.
For notation brevity, here we provide a naive graphical and attentional mechanism based encoder as the backbone for each domain.
Taking the domain $k$ as example, the \texttt{Expert}$^k$ includes:
\begin{enumerate}
    \item Graph neural network-based encoder: In recommendation, a high-level view of the entire interaction graph could reflect the steady item-item relation. Thereby, we first utilize the item relationship matrix $\mathbf{A}^k$ through a GNN to inject high-order interaction knowledge:
    \begin{equation}
    \begin{aligned}
        \mathbf{S}^k_{L}=\texttt{Norm}\left(\mathbf{A}^k\right) \mathbf{S}^k_{L-1}, \quad \text{where} \quad \mathbf{S}^k_{0} = \mathbf{S}^k,
    \end{aligned}
    \end{equation}
    where $\texttt{Norm}(\cdot)$ means row-level mean normalization, $L$ is GNN depth.
    \item Self-Attention based encoder: Instead of the graph-level modeling, the item sequence modeling is also important to make precise user preference capturing. Here we utilize causal self-attention to extract sequence representation:
    % multi-layer and multi-head
    \begin{equation}
    \begin{aligned}
        \mathbf{z}^k_T=\texttt{Causal-Self-Attention}^k
        \left((s_1^k, s_2^k,\dots,s_T^k), \mathbf{S}^k+\mathbf{S}^k_L, \mathbf{P}^{k}\right), \\
    \end{aligned}
    \end{equation}
    where $\mathbf{z}^k_T\in\mathbb{R}^d$ is representation at $T$, which aims at search next item $s_{T+1}^k$.
\end{enumerate}
Based on the graphical and attentional \texttt{Expert}$(\cdot)$, we could generate the final $\mathbf{z}^k_T$ for any item sequence $(s_1^k, s_2^k,\dots,s_T^k)$ to fit data distribution.
\begin{figure*}[t]
    \centering
    \includegraphics[width=1.0\textwidth]{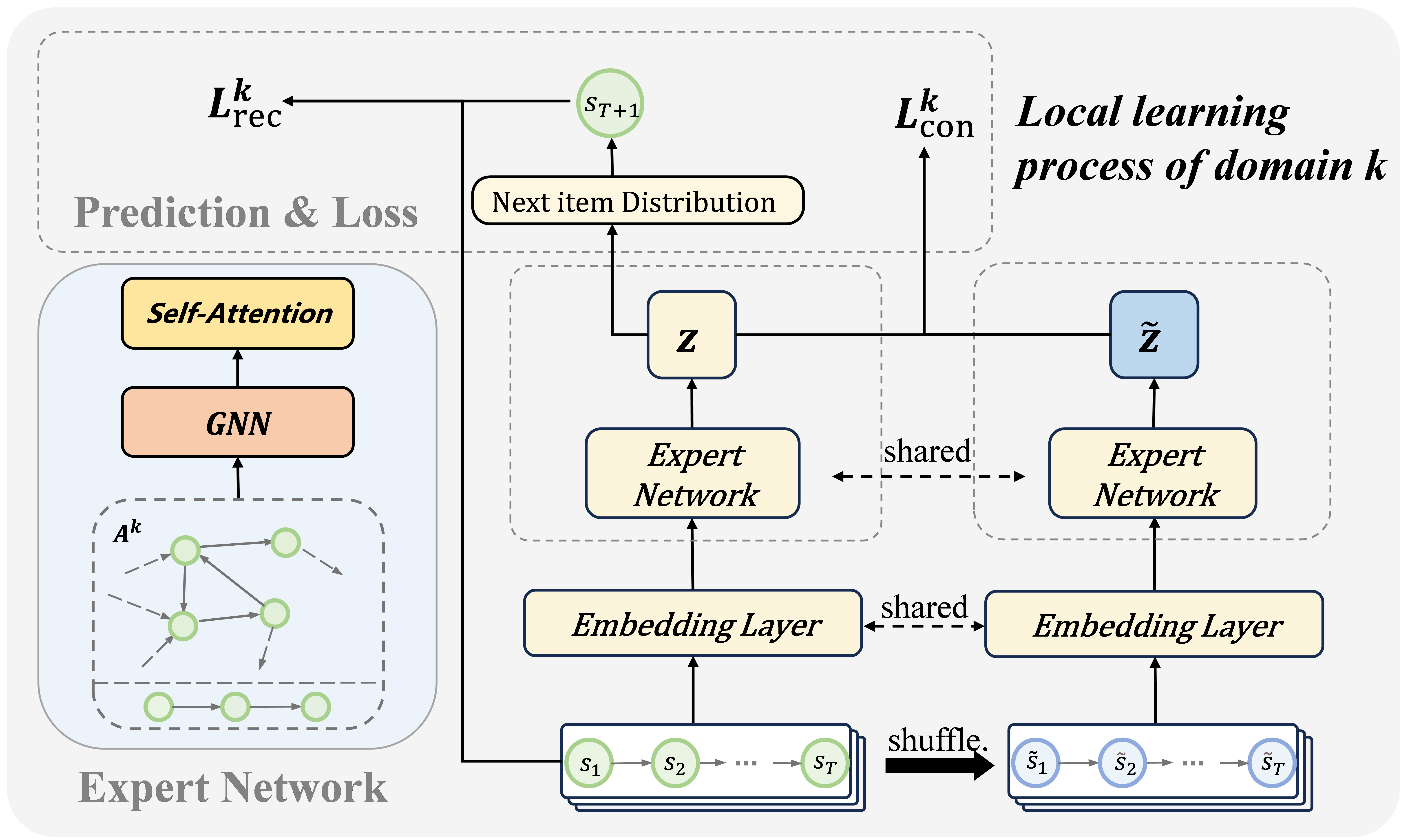}
    \caption{The local domain expert training process of domain $k$.}
    \label{Expert-framework}
\end{figure*}
Here we introduce two type of prediction tasks for robust model training:
\begin{enumerate}
    \item Augmentation contrastive objective: Unlike text in NLP, the sequence relationship in the recommendation scenario is weaker, masking or shuffling a few items will not bring about essential changes in interest significantly. To enhance the robustness of learned representation, we adopt a contrastive learning objective that the consistency between original sequence and augmented user sequences:
    \begin{equation}
        \begin{split}
            (&\tilde{s}_1^k, \tilde{s}_2^k,\dots,\tilde{s}_T^k) =  \texttt{Augment}(s_1^k, s_2^k,\dots,s_T^k) \\
            &\tilde{\mathbf{z}}^k_T=\texttt{Expert}^k((\tilde{s}_1^k, \tilde{s}_2^k,\dots,\tilde{s}_T^k), \mathbf{S}^k, \mathbf{P}^k), \\
            &\mathcal{L}_{con}^k =\texttt{In-Batch-Contrastive}(\mathbf{z}^k_T, \tilde{\mathbf{z}}^k_T)
        \end{split}
    \end{equation}
    where $\texttt{Augment}(\cdot)$ is a random shuffle function, $\texttt{In-Batch-Contrastive}(\cdot)$ is the wide used contrastive technique~\cite{cao2022contrastive}.
    \item Next item prediction objective: we adopt the commonly used training strategy to optimize our \texttt{Expert}$^k$ as follows:
        \begin{equation}
            \begin{split}
                &\mathbf{o}^k_{T+1}= \texttt{MLP}^k(\mathbf{z}_{T}^k) \\
                \mathcal{L}_{\text{rec}}^k &=  \texttt{CrossEntropy}(\mathbf{o}^k_{T+1}, s_{T+1}^k) \\
            \end{split}
        \end{equation}
        where the $\mathbf{o}^k_{T+1}\in \mathbb{R}^{|\mathcal{S}^k|}$ is the next item prediction distribution, $\texttt{MLP}(\cdot):\mathbb{R}^{d\rightarrow |\mathcal{S}^k|}$ is a multi-layer neural projector.
\end{enumerate}

\begin{figure*}[t]
    \centering
    \includegraphics[width=1.0\textwidth]{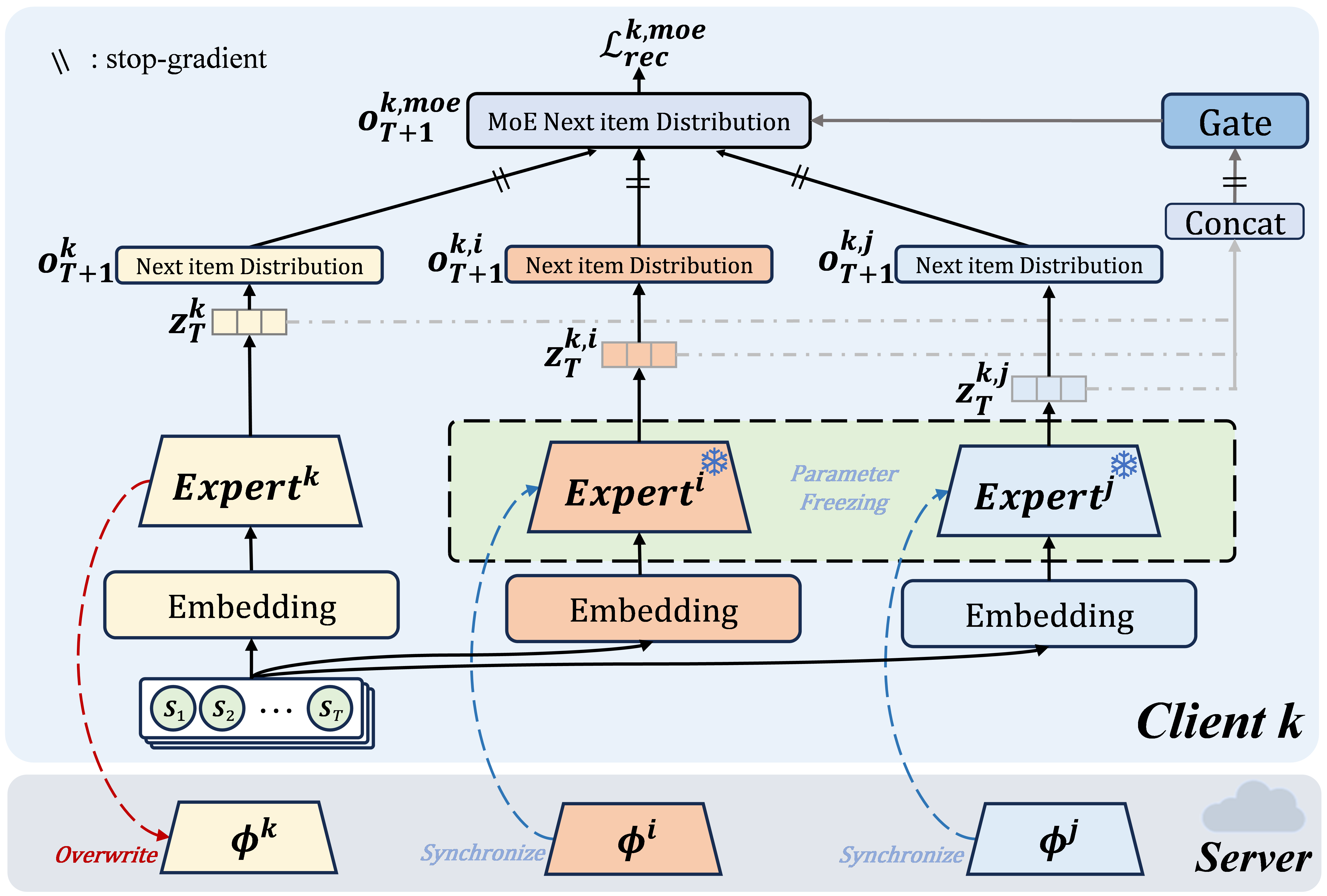}
    \caption{The Information Fusion Process of MoE.}
    \label{FedMoE-framework}
\end{figure*}

\subsection{Global Expert Adaptation}
Up to now, we have introduced how to train the local \texttt{Expert} via its in-domain data, and these local \texttt{Expert} parameters will periodically overwrite the server cache, e.g., $\texttt{Expert}^i(\cdot)\rightarrow \phi^{i}, \texttt{Expert}^j(\cdot)\rightarrow \phi^{j}, \texttt{Expert}^k(\cdot)\rightarrow \phi^{k}$.
According to server model caches of other domains, in this section, we dive into the federated paradigm to show how the global synchronized domain trained \texttt{Expert}$^i(\cdot)$, \texttt{Expert}$^j(\cdot)$ works in local domain $k$.
In practice, through periodic parameter synchronization with a central server, domain $k$ can download the latest global cached expert parameters as:
\begin{equation}
\begin{aligned}
    \texttt{Expert}^{k,i}(\cdot)\leftarrow \phi^{i}, \quad \texttt{Expert}^{k,j}(\cdot)\leftarrow \phi^{j}
\end{aligned}
\end{equation}
where the $\texttt{Expert}^{k,i}(\cdot)$ and $\texttt{Expert}^{k,j}(\cdot)$ are downloaded trained parameters on domain $k$.
In our federated paradigm, we also maintain that all global synchronized expert parameters remain \textit{frozen} during local training.
Specifically, to ensure the frozen $\texttt{Expert}^{k, i}(\cdot), \texttt{Expert}^{k,j}(\cdot)$ automatically adaptively in domain $k$, we devise a learning strategy that \textbf{gradient isolation} between different experts within each domain to ensure training stability and prevent interference.
In implement, we assign additional parameters for each global \texttt{Expert}, taking the $\texttt{Expert}^{k,i}(\cdot)$ as example:
\begin{equation}
\begin{aligned}
    &\mathbf{z}^{k,i}_T = \texttt{Expe}\texttt{rt}^{k,i}((s_1^k, s_2^k,\dots,s_T^k), \mathbf{S}^{k,i}, \mathbf{P}^{k,i}),\\
    &\tilde{\mathbf{z}}^{k,i}_T = \texttt{Expe}\texttt{rt}^{k,i}((\tilde{s}_1^k, \tilde{s}_2^k,\dots,\tilde{s}_T^k), \mathbf{S}^{k,i}, \mathbf{P}^{k,i})\\
    &\mathcal{L}_{con}^{k,i} =\texttt{In-Batch-Contrastive}(\mathbf{z}^{k,i}_T, \tilde{\mathbf{z}}^{k,i}_T)\\
    \mathcal{L}_{\text{rec}}^{k,i} = \texttt{C}&\texttt{rossEntropy}(\mathbf{o}^{k,i}_{T+1}, s_{T+1}^k), \quad \mathbf{o}^{k,i}_{T+1}= \texttt{MLP}^{k,i}(\mathbf{z}_{T}^{k,i}) \\
\end{aligned}
\end{equation}
where the $\mathbf{S}^{k,i}, \mathbf{P}^{k,i}$ and $\texttt{MLP}^{k,i}(\cdot)$ are isolation global side parameters and the $\mathcal{L}_{\text{con}}^{k,i}$/$\mathcal{L}_{\text{rec}}^{k,i}$ is the   analogous augmentation and next item prediction objective to optimize these parameters to adapt $\texttt{Expert}^{k,i}(\cdot)$.
Therefore, each global-side \texttt{Expert} has its an independent optimizing flow, which guarantees every \texttt{Expert} could maximize their effectiveness in target domain.

\subsection{Federated Mixture-of-Expert Mechanism}
On top of the gradient isolation framework trained local and global \texttt{Expert}, inspired by the mixture-of-expert, we consider building a gate router to adaptively fuse those prediction score to combine multiple domain recommendation knowledge (as shown in Figure~\ref{FedMoE-framework}).

\subsubsection{Gate Router}

Following the previous MoE efforts~\cite{ma2018modeling,wang2024home}, for each training sample, we also utilize all $\texttt{Expert}$ outputs to serve as our gate router input:
\begin{equation}
    \mathbf{z}^{k, \text{gate}}_T = \texttt{stop-gradient}(\mathbf{z}_T^{k} \Vert \mathbf{z}_T^{k,i} \Vert \mathbf{z}_T^{k,j})
\end{equation}
where \( || \) denotes concatenation along the hidden dimension and $\mathbf{z}^{gate} \in \mathbb{R}^{T \times 3d}$.
It is worth noting that we apply the \texttt{stop-gradient} operation to ensure the MoE module does not affect the learning of the each \texttt{Expert} convergency.
Next, we could predict each Expert importance as follows:
\begin{equation}
    \mathbf{g}_T^k = \texttt{Gate}^k(\mathbf{z}^{k, \text{gate}}_T)
\end{equation}
where $\texttt{Gate}^k(\cdot):\mathbb{R}^{3d\rightarrow 3}$ is a Softmax-activated neural network, $\mathbf{g}_T^k\in\mathbb{R}^3$ is expert distribution (sum is 1) for $\texttt{Expert}^k(\cdot)$, $\texttt{Expert}^{k,i}(\cdot)$ and $\texttt{Expert}^{k,j}(\cdot)$.
Particularly, the gate gradient isolation achieves dynamic integration of cross-domain patterns - the gate network can utilize the input sequence characteristics to automatically highlight relevant experts, while the global expert gradient isolation independent \texttt{Expert} preserves specialized recommendation knowledge.

\begin{algorithm2e}[t]
    \caption{FMoE-CDSR}\label{alg1}
    \KwIn{Local datasets $\mathcal{D}_i, \mathcal{D}_j, \mathcal{D}_k$, epochs $N$, local epochs $M$}
    \KwOut{The optimal encoder $\texttt{Expert}^i(\cdot),\texttt{Expert}^j(\cdot),\texttt{Expert}^k(\cdot)$ and gate network parameters $\texttt{Gate}^i(\cdot),\texttt{Gate}^j(\cdot),\texttt{Gate}^k(\cdot)$}
    % initialization\;
    \textbf{Server executes:} \\
    initialize server-side cache $\{\texttt{Expert}^{x}_0(\cdot)\}_{x\in[i, j, k]}$\;
    \For{round $t = 0, 1, \cdots N - 1$}{
    % \For{client $y \in [i, j, k]$ in parallel}{
    % send $\{\texttt{Expert}^{x,t}(\cdot)\}_{x\in[i, j, k]}$ to client $y$\;  % t=0 时发送所有参数（包括自身）
    % $\texttt{Expert}^{y,t+1}(\cdot) = \textbf{ClientUpdate}\left(y, \{\texttt{Expert}^{x,t}(\cdot)\}_{x\in[i, j, k]}\right)$\;
    $\texttt{Expert}^{i}_{t+1}(\cdot) = \textbf{ClientUpdate}\left(i\right)$\;
    $\texttt{Expert}^{j}_{t+1}(\cdot) = \textbf{ClientUpdate}\left(j\right)$\;
    $\texttt{Expert}^{k}_{t+1}(\cdot) = \textbf{ClientUpdate}\left(k\right)$\;
    % }
    }
    \SetKwFunction{FMain}{}
    \SetKwProg{Fn}{ClientUpdate}{:}{}
    \Fn{\FMain{$k$}}{
    Synchronizing global-side $\texttt{Experts}$ as $\texttt{Expert}^{k, i}$ and $\texttt{Expert}^{k, j}$.\;
    \For{local epoch $i=1, 2, \ldots, M$}{
    \For{sequence batch $(s_1^k, s_2^k,\dots,s_T^k)$ from $\mathcal{D}^k$}{
    $ (\tilde{s}_1^k, \tilde{s}_2^k,\dots,\tilde{s}_T^k) =  \texttt{Augment}(s_1^k, s_2^k,\dots,s_T^k)$\;
    % \For{$x$ in $[i, j, k]$}{
    % $\boldsymbol{z}_T^x, \widetilde{\boldsymbol{z}}_T^x = \texttt{Expert}^{y,x}\big((s_1^y, s_2^y,\dots,s_T^y); (\tilde{s}_1^y, \tilde{s}_2^y,\dots,\tilde{s}_T^y); \boldsymbol{\phi}^x\big)$\;
    % $\mathbf{o}^x_{T+1}= \texttt{MLP}^x(\mathbf{z}_{T}^x)$\;
    % $\mathcal{L}_{\text{rec}}^x =  \texttt{CrossEntropy}(\mathbf{o}^x_{T+1}, s_{T+1}^x)$\;
    % }
    $\text{Calculate }\mathcal{L}^{k}_{rec},\mathcal{L}^{k}_{con}\text{ based on }\texttt{Expert}^{k}$\;
    $\text{Calculate }\mathcal{L}^{k,i}_{rec},\mathcal{L}^{k,i}_{con} \text{ based on }\texttt{Expert}^{k,i}$\;
    $\text{Calculate }\mathcal{L}^{k,j}_{rec},\mathcal{L}^{k,j}_{con} \text{ based on }\texttt{Expert}^{k,j}$\;
    $\mathbf{z}^{k, \text{gate}}_T = \texttt{stop-gradient}(\mathbf{z}_T^{k} \Vert \mathbf{z}_T^{k,i} \Vert \mathbf{z}_T^{k,j})$\;
    $\mathbf{g}^k_T = \texttt{Gate}^k(\mathbf{z}^{k, \text{gate}}_T)$\;
    $\mathbf{o}_{T+1}^{k,moe} = \texttt{WeightedSum}\big(\{\mathbf{o}^{k}_{T+1}, \mathbf{o}^{k,i}_{T+1}, \mathbf{o}^{k,j}_{T+1}\}, \mathbf{g}_T^k\big)$\;
    $\mathcal{L}_{\text{rec}}^{k,moe} = \texttt{CrossE}\texttt{ntropy}(\mathbf{o}^{k,moe}_{T+1}, s_{T+1}^y)$\;
    % $\text{minimize}(\sum_{x=i,j,k}\mathcal{L}_{\text{rec}}^x+\mathcal{L}_{\text{rec}}^{y,moe})\;
    $ \text{Optimize all loss function } \mathcal{L}$\;
    }
    }
    \KwRet $\texttt{Expert}^k(\cdot)$;
    }
\end{algorithm2e}

\subsubsection{Knowledge Fusion}
According to the \texttt{Expert} importance results $\mathbf{g}_T^k$, and each \texttt{Expert} prediction results, e.g., $\mathbf{o}^{k}_{T+1}, \mathbf{o}^{k,i}_{T+1}, \mathbf{o}^{k,j}_{T+1}$,
the representation fusion process is defined as:
\begin{equation}
    \begin{aligned}
    \mathbf{o}_{T+1}^{k,moe} = \texttt{WeightedSum}&\big(\{\mathbf{o}^{k}_{T+1}, \mathbf{o}^{k,i}_{T+1}, \mathbf{o}^{k,j}_{T+1}\}, \mathbf{g}_T^k\big) \\
    \mathcal{L}_{\text{rec}}^{k,moe} = \texttt{CrossE}&\texttt{ntropy}(\mathbf{o}^{k,moe}_{T+1}, s_{T+1}^k)\\
    \end{aligned}
\end{equation}
where the $\mathbf{o}_{T+1}^{k,moe}$ is the fusion prediction score, and the $\mathcal{L}_{\text{rec}}^{k,moe}$ is the training objective to maximize the gate network effectiveness to select appropriate \texttt{Expert} for different training samples.
Compared with FedAvg, our federated information fusion paradigm has the following advantages:
\begin{enumerate}
    \item Effective utilization of model parameters from multiple other domains without architectural constraints.
    \item Mitigation of significant information loss to effectively reflect the unique knowledge of each domain.
    \item  Prevention of negative interference between heterogeneous domains.
\end{enumerate}

\section{Experiments}
In this section, we perform detailed experiments on three CDSR scenarios, to verify the effectiveness of our federated learning paradigm for CDSR. 
In general, we aim to answer the following research questions:
\begin{itemize}
\item \textbf{RQ1}: How does FMoE-CDSR perform with SOTA baseline and achieve satisfactory improvements?
\item \textbf{RQ2}: How does other domain model parameter quality effect the target domain performance?
\item \textbf{RQ3}: How does FMoE-CDSR components works as expected?
\item \textbf{RQ4}: How does the global model prediction accuracy at each local domain?
\end{itemize}

\subsection{Datasets}
In our study, we utilize publicly available datasets from Amazon\footnote{\url{https://jmcauley.ucsd.edu/data/amazon/}}, an open platform, to create federated cross-domain sequential recommendation (CDSR) scenarios. Following the FedDCSR data splitting, we also select ten domains to generate three multiple real-world scenarios\footnote{\url{https://drive.google.com/drive/folders/1NnZN3LhzdpxwaHiOW8GAUS8noTbdLlQt?usp=drive\_link}}: Food-Kitchen-Cloth-Beauty (FKCB), Movie-Book-Game (MBG), and Sports-Garden-Home (SGH).
Referring to previous studies, we remove users and items that have interactions fewer than ten and only retain the sequences whose number of items between four and sixteen. In the facet of dataset splitting, we select the latest 20\% of each user’s interactions as the validation/test set, while the remaining 80\% served as the training set. Table 1 summarizes the specific statistics of the federated CDSR scenarios.
\begin{table}[tbp]
\centering
\caption{Statistics of Three Federated CDSR scenarios.}
\setlength{\tabcolsep}{15pt}
\resizebox{10cm}{!}{
 \renewcommand{\arraystretch}{1.3}
\begin{tabular}{l|c|c|c|c|c|c}
\toprule
\textbf{Datasets}  & \textbf{Users} & \textbf{Items} & \textbf{Train} & \textbf{Valid} & \textbf{Test} & \textbf{Avg.Len} 
\\ \hline
Food     & 4658    & 13564   & 4977    & 1307    & 1332   & 8.79    \\
Kitchen  & 13382   & 32918   & 11100   & 2172    & 2254   & 8.60    \\
Clothing & 9240    & 34909   & 5720    & 818     & 866    & 9.30    \\
Beauty   & 5902    & 17780   & 4668    & 836     & 855    & 10.29   \\ \midrule
Movie    & 34792   & 44464   & 57405   & 10944   & 11654  & 7.97    \\
Book     & 19419   & 72246   & 63157   & 11168   & 12149  & 7.20    \\
Game     & 5588    & 10336   & 6631    & 1374    & 1444   & 6.49    \\ \midrule
Sport    & 28139   & 88992   & 51477   & 13720   & 14214  & 10.65   \\
Garden   & 6852    & 21604   & 10479   & 3074    & 3113   & 9.48    \\
Home     & 20784   & 62499   & 37361   & 10058   & 10421  & 10.41   \\ 
\bottomrule
\end{tabular}
}
\label{dataset}
\end{table}

\begin{table*}[t!]
    \large
    \centering
    \caption{Experimental results on the FKCB scenario. Avg denotes the average result calculated from all domains. The best results are boldfaced.}
    % \vspace{-.1cm}
    \resizebox{\linewidth}{!}{
        \setlength{\tabcolsep}{4pt}{
         \renewcommand{\arraystretch}{1.5}
            \begin{tabular}{l|ccc|ccc|ccc|ccc|ccc}
                \toprule
                \multirow{3}{*}{Method} & \multicolumn{3}{c|}{Food} & \multicolumn{3}{c|}{Kitchen} & \multicolumn{3}{c|}{Clothing} & \multicolumn{3}{c|}{Beauty} & \multicolumn{3}{c}{Avg}                                                                                                                                                                        \\
                \cline{2-16}            
            & \multicolumn{1}{c}{\multirow{2}{*}{MRR}} & \multicolumn{1}{c}{HR} & \multicolumn{1}{c|}{NDCG} 
            & \multicolumn{1}{c}{\multirow{2}{*}{MRR}} & \multicolumn{1}{c}{HR} & \multicolumn{1}{c|}{NDCG} 
            & \multicolumn{1}{c}{\multirow{2}{*}{MRR}} & \multicolumn{1}{c}{HR} & \multicolumn{1}{c|}{NDCG}  
            & \multicolumn{1}{c}{\multirow{2}{*}{MRR}} & \multicolumn{1}{c}{HR} & \multicolumn{1}{c|}{NDCG}  
            & \multicolumn{1}{c}{\multirow{2}{*}{MRR}} & \multicolumn{1}{c}{HR} & \multicolumn{1}{c}{NDCG}  \\
            \cline{3-4}\cline{6-7}\cline{9-10}\cline{12-13}\cline{15-16} % 只在HR和NDCG上方画线
             &  &@10 &@10
             &  &@10 &@10 
             &  &@10 &@10 
             &  &@10 &@10 
             &  &@10 &@10 \\
                \midrule
                FedSASRec               & 6.84                      & 14.41                        & 8.03                          & 0.92                        & 1.64                    & 0.95           & 0.32          & 0.46          & 0.33          & 3.66           & 7.02           & 4.14           & 2.94           & 5.88           & 3.36           \\
                \midrule
                FedVSAN                 & 21.31                     & 35.21                        & 23.92                         & 6.46                        & 12.38                   & 7.06           & 1.60          & 2.89          & 1.49          & 11.58          & 20.70          & 13.01          & 10.24          & 17.79          & 11.37          \\
                FedContrastVAE          & 23.38                     & 39.19                        & 26.53                         & 7.15                        & 12.20                   & 7.63           & 1.74          & 3.58          & 1.71          & 15.08          & 25.73          & 16.88          & 11.84          & 20.18          & 13.19          \\
                \midrule
                FedCL4SRec              & 21.89                     & 34.53                        & 24.32                         & 5.56                        & 9.80                    & 5.93           & 1.49          & 2.31          & 1.29          & 12.79          & 21.52          & 14.27          & 10.43          & 17.04          & 11.45          \\
                FedDuoRec               & 21.60                     & 33.63                        & 23.93                         & 5.45                        & 9.23                    & 5.74           & 1.61          & 2.42          & 1.38          & 13.15          & 21.05          & 14.45          & 10.45          & 16.58          & 11.38          \\
                \midrule
                FedDCSR                 & 28.87                     & 45.65                        & 32.30                         & 11.37                       & 21.96                   & 13.07          & 1.99          & 3.23          & 1.87          & 18.73          & 33.45          & 21.66          & 15.24          & 26.07          & 17.23          \\
                \midrule
                FMoE-CDSR(Ours)            & \textbf{29.97}            & \textbf{48.87}               & \textbf{33.99}                & \textbf{11.68}              & \textbf{22.54}          & \textbf{13.55} & \textbf{2.56} & \textbf{4.04} & \textbf{2.43} & \textbf{19.61} & \textbf{35.20} & \textbf{22.76} & \textbf{15.96} & \textbf{27.66} & \textbf{18.18} \\
                \bottomrule
            \end{tabular}
        }}
    %\begin{center}
    %\textbf{Boldface} and \underline{underlined} numbers denote the best and runner-up results of all methods, respectively.
    %\end{center}
    \label{main_exp}
\end{table*}
\begin{table*}[t]
    \footnotesize
    \centering
    \caption{Experimental results on the MBG scenario.}
    % \vspace{-.1cm}
     \resizebox{\linewidth}{!}{
        \setlength{\tabcolsep}{4pt}{
         \renewcommand{\arraystretch}{1.3}
            \begin{tabular}{l|ccc|ccc|ccc|ccc}
                \toprule
                \multirow{2}{*}{Method} & \multicolumn{3}{c|}{Movie} & \multicolumn{3}{c|}{Book} & \multicolumn{3}{c|}{Game}  & \multicolumn{3}{c}{Avg} 
                
                 \\
                \cline{2-13}            
            & \multicolumn{1}{c}{\multirow{2}{*}{MRR}} & \multicolumn{1}{c}{HR} & \multicolumn{1}{c|}{NDCG} 
            & \multicolumn{1}{c}{\multirow{2}{*}{MRR}} & \multicolumn{1}{c}{HR} & \multicolumn{1}{c|}{NDCG} 
            & \multicolumn{1}{c}{\multirow{2}{*}{MRR}} & \multicolumn{1}{c}{HR} & \multicolumn{1}{c|}{NDCG}  
            & \multicolumn{1}{c}{\multirow{2}{*}{MRR}} & \multicolumn{1}{c}{HR} & \multicolumn{1}{c}{NDCG}  \\
            
             \cline{3-4}\cline{6-7}\cline{9-10}\cline{12-13} % 只在HR和NDCG上方画线
             &  &@10 &@10
             &  &@10 &@10 
             &  &@10 &@10 
             &  &@10 &@10 
            \\
                \midrule
                FedSASRec               & 10.36                      & 17.79                     & 11.26                     & 6.91                    & 10.91          & 7.19           & 5.20          & 6.44           & 5.15          & 7.49           & 11.71          & 7.87           \\
                \midrule
                FedVSAN                 & 5.93                       & 12.72                     & 6.43                      & 6.54                    & 13.09          & 7.27           & 1.83          & 3.20           & 1.74          & 4.77           & 9.67           & 5.15           \\
                FedContrastVAE          & 11.25                      & 19.14                     & 12.22                     & 7.63                    & 12.76          & 8.18           & 4.95          & 6.72           & 4.96          & 7.94           & 12.87          & 8.45           \\
                \midrule
                FedCL4SRec              & 10.39                      & 17.68                     & 11.24                     & 6.86                    & 10.94          & 7.18           & 5.21          & 6.44           & 5.17          & 7.49           & 11.69          & 7.87           \\
                FedDuoRec               & 10.42                      & 17.83                     & 11.55                     & 6.79                    & 10.80          & 7.12           & 5.16          & 6.42           & 5.13          & 7.46           & 11.68          & 7.93           \\
                \midrule
                FedDCSR                 & 16.11                      & 28.32                     & 18.09                     & 10.38                   & 18.68          & 11.56          & 7.65          & 10.60          & 7.83          & 11.38          & 19.20          & 12.49          \\
                \midrule
                FMoE-CDSR(Ours)            & \textbf{18.59}             & \textbf{32.48}            & \textbf{21.00}            & \textbf{13.65}          & \textbf{23.52} & \textbf{15.15} & \textbf{8.63} & \textbf{13.23} & \textbf{9.17} & \textbf{13.62} & \textbf{23.07} & \textbf{15.10} \\
                \bottomrule
            \end{tabular}
        }}
    %\begin{center}
    %\textbf{Boldface} and \underline{underlined} numbers denote the best and runner-up results of all methods, respectively.
    %\end{center}
    \label{main_exp}
\end{table*}
\begin{table*}[t!]
    \footnotesize
    \centering
    \caption{Experimental results on the SGH scenario.}
    % \vspace{-.1cm}
     \resizebox{\linewidth}{!}{
        \setlength{\tabcolsep}{4pt}{
         \renewcommand{\arraystretch}{1.3}
            \begin{tabular}{l|ccc|ccc|ccc|ccc}
                \toprule
                \multirow{3}{*}{Method} & \multicolumn{3}{c|}{Sport} & \multicolumn{3}{c|}{Garden} & \multicolumn{3}{c|}{Home }  & \multicolumn{3}{c}{Avg} \\
                \cline{2-13}            
            & \multicolumn{1}{c}{\multirow{2}{*}{MRR}} & \multicolumn{1}{c}{HR} & \multicolumn{1}{c|}{NDCG} 
            & \multicolumn{1}{c}{\multirow{2}{*}{MRR}} & \multicolumn{1}{c}{HR} & \multicolumn{1}{c|}{NDCG} 
            & \multicolumn{1}{c}{\multirow{2}{*}{MRR}} & \multicolumn{1}{c}{HR} & \multicolumn{1}{c|}{NDCG}  
            & \multicolumn{1}{c}{\multirow{2}{*}{MRR}} & \multicolumn{1}{c}{HR} & \multicolumn{1}{c}{NDCG}  \\
            
            \cline{3-4}\cline{6-7}\cline{9-10}\cline{12-13} % 只在HR和NDCG上方画线
             &  &@10 &@10
             &  &@10 &@10 
             &  &@10 &@10 
             &  &@10 &@10 
            \\
                \midrule
                FedSASRec               & 3.59                       & 4.34                        & 3.43                      & 4.74                    & 5.36          & 4.56          & 3.13          & 3.77          & 2.93          & 3.82          & 4.49          & 3.64          \\
                \midrule
                FedVSAN                 & 1.16                       & 2.05                        & 1.00                      & 1.18                    & 2.02          & 1.05          & 1.27          & 2.26          & 1.10          & 1.21          & 2.11          & 1.05          \\
                FedContrastVAE          & 3.93                       & 4.95                        & 3.81                      & 5.07                    & 5.97          & 4.91          & 3.54          & 4.37          & 3.39          & 4.18          & 5.10          & 4.03          \\
                \midrule
                FedCL4SRec              & 3.60                       & 4.26                        & 3.41                      & 4.74                    & 5.27          & 4.55          & 3.11          & 3.74          & 2.91          & 3.82          & 4.42          & 3.62          \\
                FedDuoRec               & 3.75                       & 4.43                        & 3.64                      & 4.86                    & 5.47          & 4.79          & 3.07          & 3.60          & 2.83          & 3.89          & 4.50          & 3.75          \\
                \midrule
                FedDCSR                 & 5.29                       & 6.70                        & 5.22                      & 6.41                    & 8.35          & 6.52          & 4.52          & 5.84          & 4.45          & 5.41          & 6.96          & 5.40          \\
                \midrule
                FMoE-CDSR(Ours)            & \textbf{6.62}              & \textbf{8.20}               & \textbf{6.62}             & \textbf{7.65}           & \textbf{9.44} & \textbf{7.73} & \textbf{5.62} & \textbf{7.16} & \textbf{5.62} & \textbf{6.63} & \textbf{8.27} & \textbf{6.65} \\
                \bottomrule
            \end{tabular}
        }}
    %\begin{center}
    %\textbf{Boldface} and \underline{underlined} numbers denote the best and runner-up results of all methods, respectively.
    %\end{center}
    \label{main_exp}
\end{table*}

\subsection{Experimental Setup}
\textbf{Parameter Settings}: In our experiments, the hyperparameters are set as follows: the number of training rounds is fixed at 60 for the FKCB dataset, while it is set to 40 for the MBG and SGH datasets. The number of local epochs per client is set to 3 across all datasets. The patience value for the early stopping strategy is fixed at 10 for the FKCB dataset, while it is set to 5 for the MBG and SGH datasets. Additionally, the mini-batch size is set to 256. The learning rate is fixed at 0.001, and the dropout rate is fixed at 0.3 across all datasets.\\
\textbf{Baselines}: In this paper, we compare Ours FMoE-CDSR  with four types of representative sequential recommendation models as: (1) Attention-based methods, like SASRec~\cite{SASRec}, which first applies the self-attention network to model user sequences. (2) VAE-based methods, such as VSAN~\cite{VSAN} which utilizes variational inference to build self-attention networks for sequential recommendation, and ContrastVAE~\cite{ContrastVAE} which applies the contrastive learning to the VAE. (3) CL-based methods, including CL4SRec~\cite{CL4} and DuoRec~\cite{DuoRec} which introduces the contrastive learning framework to extract self-supervised signals from user sequences. (4) Disentangled Representation-based methods, like FedDCSR~\cite{zhang2024feddcsr} which proposes to capture intra - domain and cross - domain user preferences. We integrate (1)-(3) methods with FedAvg~\cite{fedavg} to form baselines.

\subsection{Experimental Results (RQ1)}
Tables 2, 3, and 4 show the performances of baselines and Ours FMoE-CDSR on the FKCB, MBG, and SGH CDSR scenarios. The experimental results demonstrate that our method significantly enhances
recommendation performance across multiple real-world scenarios.

From the experimental results, we have several insightful findings: (1) From the comparison between the group of VAE-based methods, we can easily find that FedContrastVAE works better than FedVSAN(about 39.92\%, 71.05\%, 13.5\% higher than FedVSAN on three scenarios), which indicates the effectiveness of adding contrastive learning into the VAE-based methods. (2) From the comparison between all the baselines, we discover that CL-based methods have better performances than attention-based methods, which indicates the excellent advantages of contrastive learning in federated CDSR scenario. (3) FedDCSR is the best-performing method aside from our method, indicating the significance of capturing intra-domain and cross-domain user preferences. (4) Besides, Our FMoE-CDSR performs best among all the baselines, proving the effectiveness of our federated mixture-of-expert paradigm and the dynamic selection of expert models attributed to the introduction of the gate network.

\begin{table}[t!]
\centering
\caption{The impact of global expert checkpoint quality on the Avg metrics on FKCB scenario.}
\setlength{\tabcolsep}{15pt}
\resizebox{12cm}{!}{
 \renewcommand{\arraystretch}{1.5}
\begin{tabular}{l|c|c|c|c}
\hline
Method  & MRR & HR@10 & NDCG@10  & Whether Local/Global Convergence
\\ \hline
Local Expert &15.25 &26.52 &17.42 &YES/-  \\ \hline
Local Expert + Global Expert(3 epochs)&13.81 &23.89 &15.61 &YES/NO    \\ \hline
Local Expert + Global Expert(5 epochs)&15.54&26.88&17.61 &YES/NO  \\ \hline
Local Expert + Global Expert(14 epochs)&\textbf{15.88}&\textbf{27.19}&\textbf{17.91}&YES/YES  \\ \hline
\end{tabular}
}
\label{quality}
\end{table}

\subsection{Ablation Study of Global Checkpoint Quality (RQ2)}
To demonstrates the effectiveness of knowledge sharing across domains, we restructure the training process into two phases: In the first phase, each local expert is trained independently; after one synchronization, we freeze the synchronized global expert and continue training until convergence without further synchronizations. From \textbf{Table~\ref{quality}}, we can conclude that higher other domain training epochs will obviously enhance the metrics of the model. This demonstrates that for a given target domain, better training on its other domain leads to improved performance after the transfer,indicating the effectiveness of knowledge sharing across domains.

\begin{table*}[t!]
    \footnotesize
    \centering
    \caption{Performance of our model variants on MBG scenario.}
    % \vspace{-.1cm}
    \resizebox{\linewidth}{!}{
        \setlength{\tabcolsep}{4pt}{
         \renewcommand{\arraystretch}{1.5}
            \begin{tabular}{l|ccc|ccc|ccc|ccc}
                \toprule
                \multirow{3}{*}{Method} & \multicolumn{3}{c|}{Movie} & \multicolumn{3}{c|}{Book} & \multicolumn{3}{c|}{Game}  & \multicolumn{3}{c}{Avg} 
                
                 \\
                \cline{2-13}            
            & \multicolumn{1}{c}{\multirow{2}{*}{MRR}} & \multicolumn{1}{c}{HR} & \multicolumn{1}{c|}{NDCG} 
            & \multicolumn{1}{c}{\multirow{2}{*}{MRR}} & \multicolumn{1}{c}{HR} & \multicolumn{1}{c|}{NDCG} 
            & \multicolumn{1}{c}{\multirow{2}{*}{MRR}} & \multicolumn{1}{c}{HR} & \multicolumn{1}{c|}{NDCG}  
            & \multicolumn{1}{c}{\multirow{2}{*}{MRR}} & \multicolumn{1}{c}{HR} & \multicolumn{1}{c}{NDCG}  \\
            
             \cline{3-4}\cline{6-7}\cline{9-10}\cline{12-13} % 只在HR和NDCG上方画线
             &  &@10 &@10
             &  &@10 &@10 
             &  &@10 &@10 
             &  &@10 &@10 
            \\
                \midrule
               
                Local Expert                  & 15.31                      & 27.30                     & 17.27                     & 10.27                   & 18.22          & 11.36          & 6.72          & 10.04          & 6.97          & 10.77          & 18.52          & 11.87          \\
                \midrule
                FMoE-CDSR -w/o Gate Router & 18.54                      & 32.17                     & 20.88                     & 13.48                   & 23.18          & 14.94          & 8.56          & 12.95          & 9.06          & 13.53          & 22.77          & 14.96          \\
                 \midrule
                FMoE-CDSR -w/o Freeze         & 16.85                      & 28.54                     & 18.76                     & 12.02                   & 20.66          & 13.32          & \textbf{9.22}          & \textbf{13.64}          & \textbf{9.71}          & 12.70          & 20.95          & 13.93          \\
                \midrule
                FMoE-CDSR(Ours)               & \textbf{18.59}             & \textbf{32.48}            & \textbf{21.00}            & \textbf{13.65}          & \textbf{23.52} & \textbf{15.15} &8.63 &13.23 &9.17 & \textbf{13.62} & \textbf{23.07} & \textbf{15.10} \\
                \bottomrule
            \end{tabular}
        }}
    %\begin{center}
    %\textbf{Boldface} and \underline{underlined} numbers denote the best and runner-up results of all methods, respectively.
    %\end{center}
    \label{abl_mbg}
\end{table*}
\begin{table*}[t!]
    \footnotesize
    \centering
    \caption{Performance of our model variants on SGH scenario.}
    % \vspace{-.1cm}
    \resizebox{\linewidth}{!}{
        \setlength{\tabcolsep}{4pt}{
         \renewcommand{\arraystretch}{1.5}
            \begin{tabular}{l|ccc|ccc|ccc|ccc}
                \toprule
                \multirow{3}{*}{Method} & \multicolumn{3}{c|}{Sport} & \multicolumn{3}{c|}{Garden} & \multicolumn{3}{c|}{Home}  & \multicolumn{3}{c}{Avg} 
                
                 \\
                \cline{2-13}            
            & \multicolumn{1}{c}{\multirow{2}{*}{MRR}} & \multicolumn{1}{c}{HR} & \multicolumn{1}{c|}{NDCG} 
            & \multicolumn{1}{c}{\multirow{2}{*}{MRR}} & \multicolumn{1}{c}{HR} & \multicolumn{1}{c|}{NDCG} 
            & \multicolumn{1}{c}{\multirow{2}{*}{MRR}} & \multicolumn{1}{c}{HR} & \multicolumn{1}{c|}{NDCG}  
            & \multicolumn{1}{c}{\multirow{2}{*}{MRR}} & \multicolumn{1}{c}{HR} & \multicolumn{1}{c}{NDCG}  \\
            
             \cline{3-4}\cline{6-7}\cline{9-10}\cline{12-13} % 只在HR和NDCG上方画线
             &  &@10 &@10
             &  &@10 &@10 
             &  &@10 &@10 
             &  &@10 &@10 
            \\
                \midrule
               
                Local Expert                  & 5.08                       & 6.55                      & 5.04                      & 6.00                    & 7.74           & 6.05           & 4.29          & 5.62           & 4.20          & 5.12           & 6.64           & 5.10           \\
                \midrule
                FMoE-CDSR -w/o Gate Router & 6.56                       & \textbf{8.24}             & 6.57                      & 7.34                    & \textbf{9.48}  & 7.48           & 5.65          & 7.10           & 5.61          & 6.52           & 8.27           & 6.55          \\
                 \midrule
                FMoE-CDSR -w/o Freeze         & 6.40                       & 7.96                      & 6.39                      & 7.49                    & 9.38           & 7.57           & \textbf{5.66} & 7.05           & 5.62          & 6.52           & 8.13           & 6.53          \\
                \midrule
                FMoE-CDSR(Ours)               & \textbf{6.62}              & 8.20                      & \textbf{6.62}             & \textbf{7.65}           & 9.44           & \textbf{7.73}  & 5.62          & \textbf{7.16}  & \textbf{5.62} & \textbf{6.63}  & \textbf{8.27} & \textbf{6.65} \\
                \bottomrule
            \end{tabular}
        }}
    %\begin{center}
    %\textbf{Boldface} and \underline{underlined} numbers denote the best and runner-up results of all methods, respectively.
    %\end{center}
    \label{abl_sgh}
\end{table*}

\subsection{Ablation Study of FMoE-CDSR Variants (RQ3)}
To investigate the effectiveness of our model components, we implement several variants of FMoE-CDSR and conduct the experiment on MBG and SGH scenarios for the task, shown in \textbf{Table~\ref{abl_mbg} and Table~\ref{abl_sgh}}.
\begin{itemize}
\item Local Expert. To verify whether our proposed federated method is useful to enhance the effectiveness of the prediction function, we first remove it. The experimental results are reported in Table~\ref{abl_mbg} and Table~\ref{abl_sgh}.The poor performance(about 16\% lower than Ours FMoE-CDSR) of Local Expert, without the federated method, shows the importance of federated learning in cross-domain recommendations. 
\item FMoE-CDSR -w/o Gate Router. FMoE-CDSR - w/o Gate Router is the MoE model without gate-weight, aggregating the predicted results using the same weights. We conduct this experiment to verify the significance of gate-weight. As expected, the experimental results meet our hypothesis. We can observe that ours FMoE-CDSR performs better than it. But compared with Local Expert, gate-weight's influence obviously lighter than federated learning. We also observe that even after removing the gate-weight strategy, the effect brought by federated aggregation is still significantly better than that of local models.
\item FMoE-CDSR -w/o Freeze. FMoE-CDSR -w/o Freeze is the FMoE-CDSR model that no longer freezes global experts parameters while training the local domain model. From the result, we can discover the \textit{freeze} of synchornized global expert parameters plays a crucial role in the prevention of negative interference between heterogeneous domains. And its outperforming compared with Local Expert verifies the effectiveness of federated aggregation again.
\end{itemize}

\begin{figure*}[t!]
    \centering
    \includegraphics[width=1.0\textwidth]{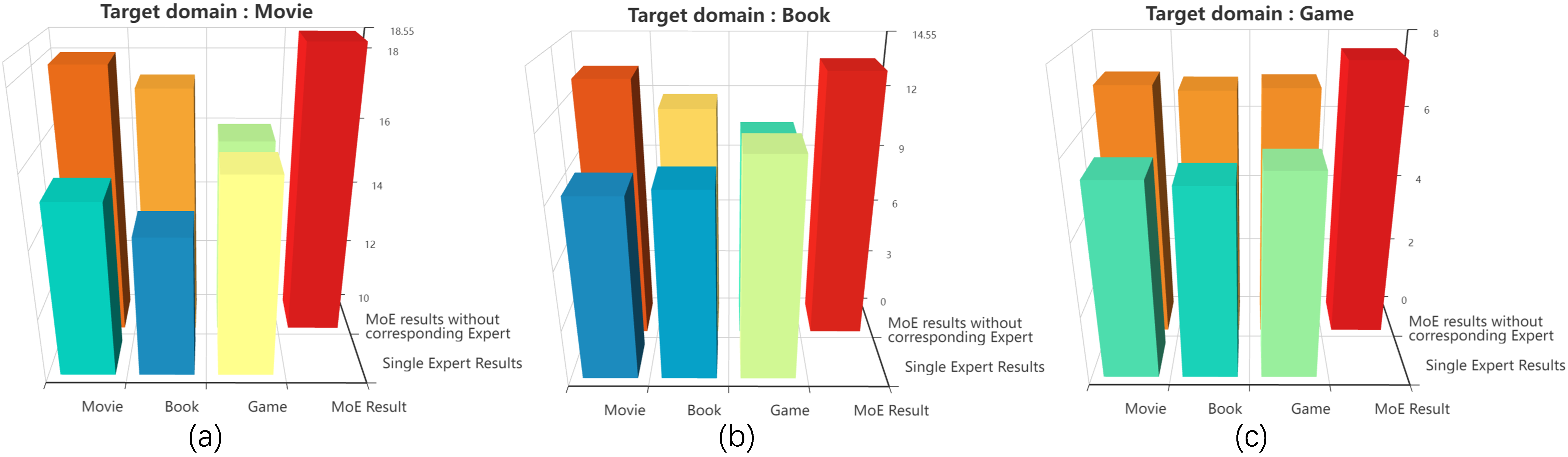}
    \caption{The predictive results of experts in MBG scenario.}
    \label{MBG_3d}
\end{figure*}
\begin{figure*}[t!]
    \centering
    \includegraphics[width=1.0\textwidth]{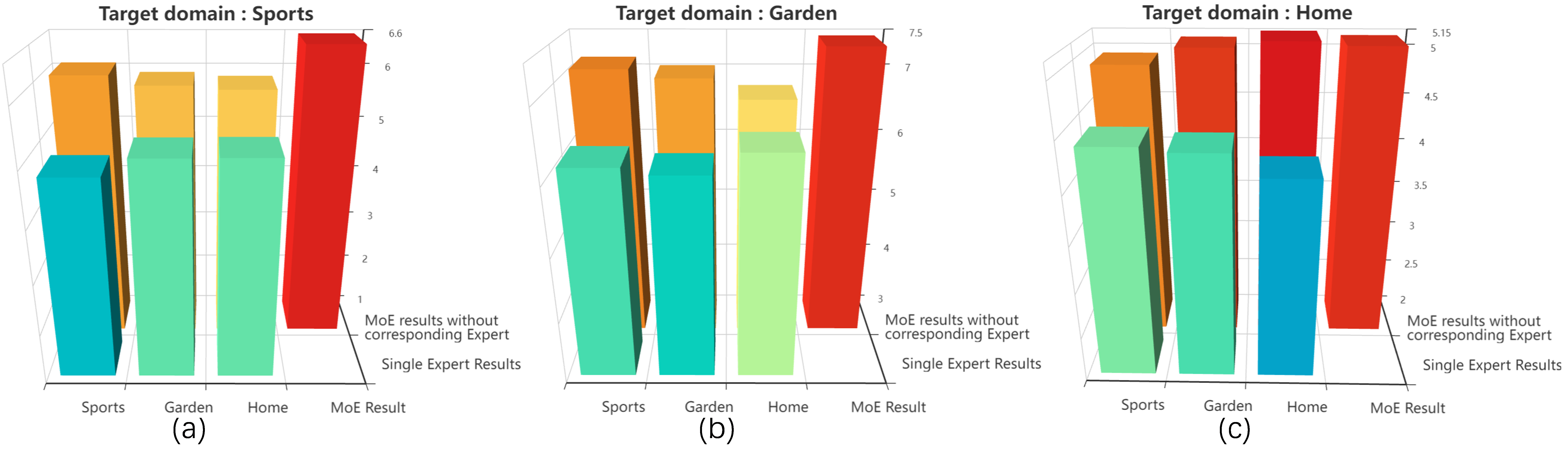}
    \caption{The predictive results of experts in SGH scenario.}
    \label{SGH_3d}
\end{figure*}

\subsection{Ablation Study of Removing Some Domains (RQ4)}
To prove the effective feature's capturing and utilizing potential across domains of ours model, we conduct the experiments by removing the other domain on the target MRR. Based on the 3D image in \textbf{Figures \ref{MBG_3d} and \ref{SGH_3d}}, we can clearly observe that the first row of cylinders represents the performance of each other domain expert when individually transferred to the target domain. For example, we assess how the Movies domain expert performs after being transferred to the Books domain, as well as the results for the Games domain expert. The first row presents the metrics of each individual expert in the target domain. The second row illustrates the performance in the target domain after removing a specific expert while aggregating all the other domain experts within the model. The figure shows that removing any migrated domain, model (e.g., Target domain: Games, remove Movies: -7.77\%, remove Books: -9.49\%, remove Games: -8.69\%) consistently degrades
performance, proving the necessity of multi-domain knowledge integration. The
negative percentage changes highlight each domain's unique contribution,
validating that FMoE-CDSR's learnable embeddings and MoE weighting effectively
utilizes complementary features across domains.

\section{Related Work}
Here we review the prior work based on differential privacy~\cite{li2023prispark} and FedAvg~\cite{fedavg}.
\textbf{Methods based on differential privacy:}
FedCDR~\cite{fedcdr} is a federated learning method for cross-domain recommendations. It relies on overlapped users and uses local differential privacy to secure embedding transfers.
CCMF~\cite{ccmf} and PriCDR~\cite{PriCDR} add noise to the original user-item matrix using differential privacy to protect privacy. They then share this modified matrix directly with the target domain.
ADPMF~\cite{ADPMF} introduces a privacy-preserving recommendation method that applies differential privacy to matrix factorization.
These studies assume overlapped users exist across different domains. They focus on transferring user embeddings via differential privacy for cross-domain knowledge transfer. However, in real-world scenarios, user overlap is often limited, which restricts the generalization ability of these methods.
To address the issue,some methods have made improvements.
CoFedRec~\cite{CoFedRec} groups users based on item categories and incorporates global item relationships in local training, avoiding the need for overlapping user assumptions.
FedUD~\cite{fedud} targets cross-platform federated click-through rate prediction using both aligned and unaligned data. Its innovation is a knowledge distillation module that transfers insights from aligned data to unaligned data, eliminating the need for overlapped users.

% \item
\textbf{Methods based on FedAvg:}
%\item 
DP-FedAvg~\cite{dpfedavg} enhances the FedAvg algorithm with noise addition and clipping to ensure user-level privacy.
P2FCDR~\cite{P2FCDR} introduces a dual-objective cross-domain recommendation framework using federated learning and local differential privacy. It improves performance by fusing features while preserving data privacy.
FedDCSR~\cite{zhang2024feddcsr} introduces a federated decoupled representation learning framework. It improves feature heterogeneity in cross-domain sequential data while protecting user privacy.
These studies use FedAvg to aggregate client model parameters, which requires identical model structures. However, user behavior varies widely across domains. Simple averaging therefore loses key domain-specific knowledge and misses unique patterns.
To address the issue,some methods have made improvements.
DP-FedEmb~\cite{DP-FedEmb} introduces a user-level differential privacy framework. It reduces reliance on homogeneous models in traditional DP-FedAvg~\cite{DP-FedAvg} by using virtual client grouping, partial parameter aggregation, and local fine-tuning.

\section{Conclusion}

In this paper, we proposed FMoE-CDSR, a federated learning framework for non-overlapped cross-domain sequential recommendation.
Unlike traditional cross-domain methods that require sharing user historical logs and embeddings, our approach improves prediction accuracy while preserving user privacy.
To achieve this, we introduced the Mixture-of-Experts (MoE)  paradigm, which effectively transfers knowledge from multiple other domains by using model parameter checkpoints. 
This mechanism enhances the adaptability and effectiveness of other domain experts in the target domain.
Extensive experiments on real-world datasets demonstrate that FMoE-CDSR significantly improves recommendation performance.
Additionally, detailed ablation analyses confirm the effectiveness of our model learning paradigm.
In the future, we plan to explore more efficient aggregation strategies and extend our framework to broader recommendation scenarios.

\bibliographystyle{splncs04}
\bibliography{mybibliography}

\begin{thebibliography}{10}
\providecommand{\url}[1]{\texttt{#1}}
\providecommand{\urlprefix}{URL }
\providecommand{\doi}[1]{https://doi.org/#1}

\bibitem{ADPMF}
Berlioz, A., Friedman, A., Kaafar, M.A., Boreli, R., Berkovsky, S.: Applying differential privacy to matrix factorization. In: ACM Conference on Recommender Systems (RecSys) (2015)

\bibitem{cao2022contrastive}
Cao, J., Cong, X., Sheng, J., Liu, T., Wang, B.: Contrastive cross-domain sequential recommendation. In: ACM International Conference on Information and Knowledge Management (CIKM). pp. 138--147 (2022)

\bibitem{cao2023towards}
Cao, J., Li, S., Yu, B., Guo, X., Liu, T., Wang, B.: Towards universal cross-domain recommendation. In: ACM International Conference on Web Search and Data Mining (WSDM) (2023)

\bibitem{cao2022disencdr}
Cao, J., Lin, X., Cong, X., Ya, J., Liu, T., Wang, B.: Disencdr: Learning disentangled representations for cross-domain recommendation. In: International ACM SIGIR Conference on Research and Development in Information Retrieval (SIGIR) (2022)

\bibitem{cao2024moment}
Cao, J., Wang, S., Li, Y., Wang, S., Tang, J., Wang, S., Yang, S., Liu, Z., Zhou, G.: Moment\&cross: Next-generation real-time cross-domain ctr prediction for live-streaming recommendation at kuaishou. ArXiv  (2024)

\bibitem{PriCDR}
Chen, C., Wu, H., Su, J., Lyu, L., Zheng, X., Wang, L.: Differential private knowledge transfer for privacy-preserving cross-domain recommendation. In: The ACM web conference (WWW) (2022)

\bibitem{chen2024multi}
Chen, G., Sun, R., Jiang, Y., Cao, J., Zhang, Q., Lin, J., Li, H., Gai, K., Zhang, X.: A multi-modal modeling framework for cold-start short-video recommendation. In: ACM Conference on Recommender Systems (RecSys) (2024)

\bibitem{P2FCDR}
Chen, G., Zhang, X., Su, Y., Lai, Y., Xiang, J., Zhang, J., Zheng, Y.: Win-win: a privacy-preserving federated framework for dual-target cross-domain recommendation. In: AAAI Conference on Artificial Intelligence (AAAI) (2023)

\bibitem{DP-FedAvg}
Cheng, A., Wang, P., Zhang, X.S., Cheng, J.: Differentially private federated learning with local regularization and sparsification. In: IEEE/CVF Computer Vision and Pattern Recognition Conference(CVPR) (2022)

\bibitem{cheng2025choruscvr}
Cheng, W., Lu, Y., Xia, B., Cao, J., Xu, K., Wen, M., Jiang, W., Zhang, J., Liu, Z., Gai, K., et~al.: Choruscvr: Chorus supervision for entire space post-click conversion rate modeling. ArXiv  (2025)

\bibitem{ccmf}
Gao, C., Huang, C., Yu, Y., Wang, H., Li, Y., Jin, D.: Privacy-preserving cross-domain location recommendation. The ACM on Interactive, Mobile, Wearable and Ubiquitous Technologies  (2019)

\bibitem{CoFedRec}
He, X., Liu, S., Keung, J., He, J.: Co-clustering for federated recommender system. In: The ACM Web Conference (WWW) (2024)

\bibitem{kairouz2021advances}
Kairouz, P., McMahan, H.B., Avent, B., Bellet, A., Bennis, M., Bhagoji, A.N., Bonawitz, K., Charles, Z., Cormode, G., Cummings, R., et~al.: Advances and open problems in federated learning. Foundations and trends{\textregistered} in machine learning  (2021)

\bibitem{SASRec}
Kang, W.C., McAuley, J.: Self-attentive sequential recommendation. In: IEEE International Conference on Data Mining (ICDM) (2018)

\bibitem{li2023prispark}
Li, S., Wen, Y., Xue, T., Wang, Z., Wu, Y., Meng, D.: Prispark: Differential privacy enforcement for big data computing in apache spark. In: International Symposium on Reliable Distributed Systems (SRDS) (2023)

\bibitem{ma2018modeling}
Ma, J., Zhao, Z., Yi, X., Chen, J., Hong, L., Chi, E.H.: Modeling task relationships in multi-task learning with multi-gate mixture-of-experts. In: ACM SIGKDD international conference on knowledge discovery \& data mining (KDD) (2018)

\bibitem{fedavg}
McMahan, B., Moore, E., Ramage, D., Hampson, S., y~Arcas, B.A.: Communication-efficient learning of deep networks from decentralized data. In: International Conference on Artificial Intelligence and Statistics (AISTATS) (2017)

\bibitem{dpfedavg}
McMahan, H.B., Ramage, D., Talwar, K., Zhang, L.: Learning differentially private recurrent language models. ArXiv  (2017)

\bibitem{fedud}
Ouyang, W., Dong, R., Tao, R., Liu, X.: Fedud: exploiting unaligned data for cross-platform federated click-through rate prediction. In: International ACM SIGIR Conference on Research and Development in Information Retrieval (SIGIR) (2024)

\bibitem{DuoRec}
Qiu, R., Huang, Z., Yin, H., Wang, Z.: Contrastive learning for representation degeneration problem in sequential recommendation. In: ACM International Conference on Web Search and Data Mining (WSDM) (2022)

\bibitem{wang2024home}
Wang, X., Cao, J., Fu, Z., Gai, K., Zhou, G.: Home: Hierarchy of multi-gate experts for multi-task learning at kuaishou. ArXiv  (2024)

\bibitem{ContrastVAE}
Wang, Y., Zhang, H., Liu, Z., Yang, L., Yu, P.S.: Contrastvae: Contrastive variational autoencoder for sequential recommendation. In: ACM International Conference on Information and Knowledge Management (CIKM) (2022)

\bibitem{CL4}
Xie, X., Sun, F., Liu, Z., Wu, S., Gao, J., Zhang, J., Ding, B., Cui, B.: Contrastive learning for sequential recommendation. In: IEEE International Conference on Data Engineering (ICDE) (2022)

\bibitem{DP-FedEmb}
Xu, Z., Collins, M., Wang, Y., Panait, L., Oh, S., Augenstein, S., Liu, T., Schroff, F., McMahan, H.B.: Learning to generate image embeddings with user-level differential privacy (2023)

\bibitem{fedcdr}
Yan, D., Zhao, Y., Yang, Z., Jin, Y., Zhang, Y.: Fedcdr: Privacy-preserving federated cross-domain recommendation. Digital Communications and Networks  (2022)

\bibitem{zang2022survey}
Zang, T., Zhu, Y., Liu, H., Zhang, R., Yu, J.: A survey on cross-domain recommendation: taxonomies, methods, and future directions. ACM Transactions on Information Systems (TOIS)  (2022)

\bibitem{zhang2024feddcsr}
Zhang, H., Zheng, D., Yang, X., Feng, J., Liao, Q.: Feddcsr: Federated cross-domain sequential recommendation via disentangled representation learning. In: SIAM International Conference on Data Mining (SDM) (2024)

\bibitem{VSAN}
Zhao, J., Zhao, P., Zhao, L., Liu, Y., Sheng, V.S., Zhou, X.: Variational self-attention network for sequential recommendation. In: IEEE International Conference on Data Engineering (ICDE) (2021)

\end{thebibliography}

\end{document}